\documentclass{article}
\usepackage{lmodern}
\usepackage{amssymb,amsmath}
\usepackage{ifxetex,ifluatex}
\usepackage{fixltx2e} %
\ifnum 0\ifxetex 1\fi\ifluatex 1\fi=0 %
  \usepackage[T1]{fontenc}
  \usepackage[utf8]{inputenc}
\else %
  \ifxetex
    \usepackage{mathspec}
  \else
    \usepackage{fontspec}
  \fi
  \defaultfontfeatures{Ligatures=TeX,Scale=MatchLowercase}
\fi
\newcommand{\euro}{EUR}
\IfFileExists{upquote.sty}{\usepackage{upquote}}{}
\IfFileExists{microtype.sty}{%
\usepackage[]{microtype}
\UseMicrotypeSet[protrusion]{basicmath} %
}{}
\PassOptionsToPackage{hyphens}{url} %
\usepackage[unicode=true]{hyperref}
\hypersetup{
            pdfborder={0 0 0},
            breaklinks=true}
\usepackage{longtable,booktabs}
\IfFileExists{footnote.sty}{\usepackage{footnote}\makesavenoteenv{long table}}{}
\usepackage{graphicx,grffile}
\makeatletter
\def\maxwidth{\ifdim\Gin@nat@width>\linewidth\linewidth\else\Gin@nat@width\fi}
\def\maxheight{\ifdim\Gin@nat@height>\textheight\textheight\else\Gin@nat@height\fi}
\makeatother
\setkeys{Gin}{width=\maxwidth,height=\maxheight,keepaspectratio}
\IfFileExists{parskip.sty}{%
\usepackage{parskip}
}{%
\setlength{\parindent}{0pt}
\setlength{\parskip}{6pt plus 2pt minus 1pt}
}
\ifx\paragraph\undefined\else
\let\oldparagraph\paragraph
\renewcommand{\paragraph}[1]{\oldparagraph{#1}\mbox{}}
\fi
\ifx\subparagraph\undefined\else
\let\oldsubparagraph\subparagraph
\renewcommand{\subparagraph}[1]{\oldsubparagraph{#1}\mbox{}}
\fi

\makeatletter
\def\fps@figure{htbp}
\makeatother

\begin{document}

\date{}
\title{A decentralized method for making sensor measurements tamper-proof to support open science applications}
\author{Patrick Wortner \and Moritz Schubotz \and Corinna Breitinger \and Stephan Leible \and Bela Gipp \and
        \normalsize School of Electrical, Information and Media Engineering\\
        \normalsize University of Wuppertal\\
        \normalsize Wuppertal Germany\\
        \normalsize \texttt{lastname@uni-wuppertal.de}}
\maketitle

\begin{abstract}
Open science has become a synonym for modern, digital and inclusive
science. Inclusion does not stop at open access. Inclusion also requires
transparency through open datasets and the right and ability to take
part in the knowledge creation process. This implies new challenges for
digital libraries. Citizens should be able to contribute data in a
curatable form to advance science. At the same time, this data should be
verifiable and attributable to its owner. Our research project focusses
on securing and attributing incoming data streams from sensors. Our
contribution is twofold. First, we analyze the promises of open science
measurement data and point out how Blockchain technology changed the
circumstances for data measurement in science projects using sensors.
Second, we present an open hardware project capable of securing the
integrity of data directly from the source using cryptographic methods.
By using inexpensive modular components and open source software, we
lower the barrier for participation in open science projects. We show
how time series of measurement values using sensors, e.g., temperature,
current, and vibration measurements, can be verifiably and immutably
stored. The approach we propose enables time series data to be stored in
a tamper-proof manner and securely timestamped on a blockchain to
prevent any subsequent modification.
\end{abstract}

CCS CONCEPTS

- Information systems $\to$ Distributed storage

KEYWORDS

Open Science, Blockchain, Data Storage, Sensor Data

\begin{figure}
\centering
\includegraphics[width=.9\textwidth]{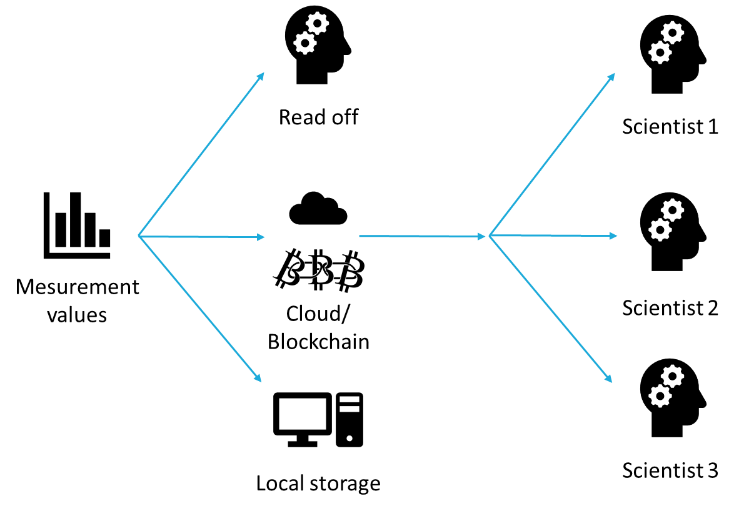}
\caption{The proposed vison of storing scientific measurements in a decentralized and trusted manner}
\end{figure}
\section{Introduction}

Traditionally, the workflow of research scientist has typically
consisted of the following steps: (1) researching literature, (2)
formulating hypothesis, (3) performing measurements, (4) evaluating
data, and (5) publishing the results. If the fifth step is repeatedly
unsuccessful, scientists in today's research environment cannot pursue a
successful career. In part due to this pressure to successfully publish,
there are regular occurrences of scientists only reporting positive
experimental outcomes, or pruning their data -- i.e., attempting to
manipulate steps 3 and 4 of the research cycle -- in order to publish
more successfully {[}5, 6{]}. In contrast to traditional science, which
has typically been performed behind closed doors, today's open-science
movement has encouraged and facilitated the publishing of measurements
(Goal A of open science). In particular, Goal A has introduced new
challenges for digital libraries.

Performing research according to open science principles {[}12{]}
provides at least two significant advantages. First, retrospective data
pruning or manipulation can become detectable, and second, other
researchers may be able to obtain additional results and reuse datasets
for new purposes without the time-consuming measurement step. Another
component of the open-science movement is a blurring of the barrier
between scientists and citizens by enabling an inclusion of any
interested community in so-called `citizen science projects' (Goal B of
open science).

Currently, there is no agreement among scientists on any standardized
procedures for reviewing openly available datasets in a way that could
be analogous to today's literature-assessment process. Methods for
researching and reviewing scientific literature by combining
content-based assessments and bibliometric measures are well established
for traditional publications. While bibliometric measures could also be
applied to datasets when these are citable, a content-based assessment
based on the raw data hardly seems possible. In contrast to the
authenticity of physical analytes which can be assayed using forensic
methods, digital datasets are of a fundamentally different nature and
their quality and integrity cannot be assessed by other researchers post
creation.

In contrast, digital measurements of real-world phenomena are a snapshot
of a state at a particular time, which is as accurate as the capturing
technology at that given moment. After the information is digitized,
there is no way to capture additional information that could be used by
future technology to evaluate the authenticity of the data. Therefore,
scientists could benefit from a method to ensure the trustworthiness of
sensor data immediately as it is generated by a measurement device
(\textbf{challenge I}).

Regarding challenge I, the law of large numbers suggests that if the
same experiments are repeated over and again, the average results will
be close to the expected outcome. Thus, by distributing the measurements
in a citizen science fashion (goal B) to different parties' errors due
to measurement equipment malfunction, and even fraudulent intentions (as
long as they do not collude) will average out. Systematic data
manipulation becomes more difficult, since multiple independent players
would need to be corrupt.

The trustworthiness of data measurements concerns the following aspects:
(1) well-defined error estimates, (2) accurate meta-data, (3) high
redundancy, (4) adequate sampling rate, (5) confirmation of all
currently known physical laws. At the same time, researchers would
benefit from a method to guarantee the immutability and durability of
time-series research data (\textbf{challenge II}). We suggest that
challenge II could be addresses by a technical solution that enables
data to be protected against subsequent manipulation through
cryptographic hashing and blockchain technology (BT), including
decentralized trusted timestamping using blockchain {[}8{]}. The
tamper-proof nature of BT could be used to, for the first time, ensure
the permanent immutability of time-series data in the digital era.

In this paper, we make the following contributions: We enumerate the
challenges to ensure the trustworthiness of time series of
measurement-data (challenge I). Moreover, we present a simple and
inexpensive open hardware solution that can be used by citizens to
measure physical properties, such as temperature, electric current, and
vibration, and an open source software that guarantees the immutability
of the measured values using Bitcoin's blockchain (challenge II).

The remaining paper is structed as follows. In Section 2, we review
related work. In Section 3, we describe our experimental setup, before
enumerating challenges and discussing future steps in Section 4.

\section{Combining citizen science and blockchain technology}

The BT underlying cryptocurrencies, such as Bitcoin {[}11{]}, offers a
bulk of characteristics, for example, decentralization, immutability,
and decentralized trusted timestamping {[}14, 17{]}, which make it
unique and have enabled both research and industry to develop new
solutions for a variety of use cases {[}3{]}.

BT has found rapid adoption to support not only financial and industry
application, but also academia {[}9{]}. Several use cases have been
proposed to support scientists and academics, such as managing academic
reputation {[}15{]}, protecting the intellectual property of academic
manuscripts submitted for peer review {[}7{]}, or tracking the
individual contributions in collaborative research projects {[}13{]}. In
this paper, we propose using blockchain technology as a foundation to
create a secure and reliable method to perform citizen science projects.
To achieve this aim, we also propose to integrate the previously
proposed concept of decentralized trusted timestamping {[}8{]}.

A blockchain can be defined as a decentralized database without a
central authority in charge of managing the data stored on it {[}1{]}.
All data stored on a blockchain are by design immutable. We use this
tamper-proof characteristic of blockchains to ensure that each
measurement value recorded by a sensor can be made immutable to
subsequent manipulation or deletion, as well as traceable to its owner.
With this contribution, we hope to further support the viability of
citizen science projects and today's open science movement by making
data and entire datasets more trustworthy.

The validity or accuracy of collected measurement data can, of course,
never be automatically guaranteed. However, in the case of citizen
science projects, the effort to manipulate a significant number of
sensors in a decentralized network would not justify the cost-benefit
factor and is additionally very unlikely. The easiest methods of
manipulation in crowd-sourced data collection experiments would be to
adjust data \emph{after} it has been measured and aggregated, which we
prevent that with our approach.

We do not propose to store all raw measurement data directly on a
blockchain, instead, we only store hashes of the data, which is more
efficient in the ways of performance, cost, and scalability to prove
that the data existed in a certain format at a certain time, thus
increasing transparency and trustworthiness.

\begin{figure}
\centering
\includegraphics[width=.5\textwidth]{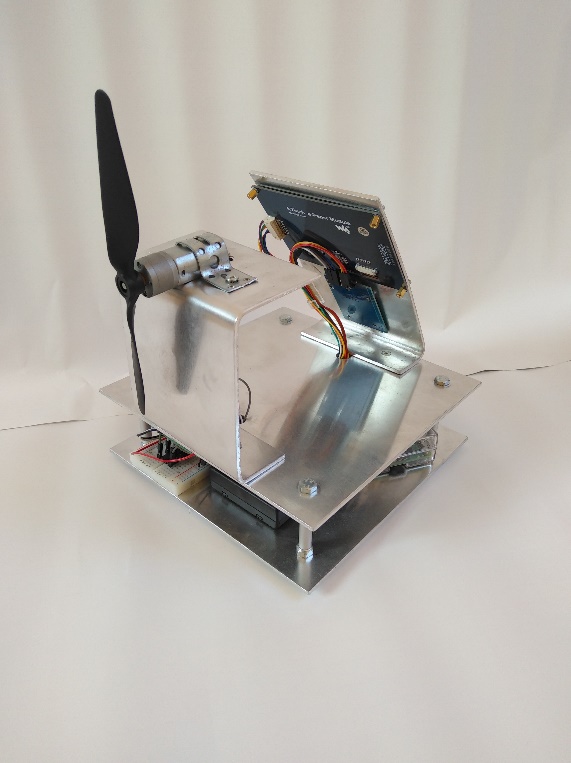}
\caption{Open hardware prototype consisting of basic and testing module to demonstrate commonly used sensors and their simple integration into our proposed system.}
\end{figure}

Once a hash has been stored on a blockchain it can always be proved that
the data (for example, sensor measurement values) were not manipulated
after they were collected {[}4{]}. From a technical point of view, there
is much literature that explains blockchain technology, how it works and
its strengths and weaknesses in different stress tests and use cases
{[}14, 16, 17{]}. Thus, we only described the special features of the
technology that we exploit for our use case. The idea is to use open
source software and hardware in combination with BT, which we implemented in an example
measurement device set-up that we will present in the next section. The
benefits in combining open science experiments with BT are the creation
of a reliable trustworthy infrastructure for building digital libraries
consisting of measurement data with superior data security and
integrity.

\section{Bootstrapping Citizen Science Projects}

Combining BT with citizen science to capture time series of sensor data,
practically means building an interface between the physical world and a
particular blockchain. In general, this would require advanced skills in
electrical engineering and blockchain programming. To lower this
entrance barrier, we developed a simple and inexpensive open hardware
prototype that captures vibration, electric current, and temperature
measurements. Our step-by-step instructions can also guide people with a
limited knowledge of single board computers and blockchain technology
through this process. Upon completion of our tutorial, the modular
design of the open hardware allows for easy customization to specific
project needs.

For our basic module, we used a Raspberry Pi model 3, a general-purpose
input/output board and an initial set of three sensors (cf. Table 1). To
test the sensors of the model, we designed a testing module - a rotor
with a battery power supply. The rotor from the testing module consumes
power and generates vibrations, which generates signals that we can
measure with the sensors from the basic module. Detailed instructions on
the building process are available from the project's GitHub repository:

\url{https://github.com/FellowsFreiesWissen/Blockchain\_Pi}

Figure 2 displays the first instance of our open hardware project (basic
and testing module). The components of our hardware setup are very
affordable (cf. list of components, Table 1), and sums only to
approximately 71 euros. The building time for the hardware is
approximately one hour.

\begin{table}
\caption{Hardware components and prices}

\begin{longtable}[]{@{}ll@{}}
\toprule
\textbf{Type} & \textbf{Price approx. in \euro{}}\tabularnewline
\midrule
\endhead
Raspberry Pi 3 & 30\tabularnewline
Case & 10\tabularnewline
GPIO Breakout and Wires & 8\tabularnewline
Vibration sensor SW-420 & 1\tabularnewline
Current sensor WCS1800 & 7\tabularnewline
Temperature sensor DHT11 & 5\tabularnewline
DC Engine (optional) & 5\tabularnewline
Battery pack (optional) & 5\tabularnewline
\textbf{Total} & \textbf{71}\tabularnewline
\bottomrule
\end{longtable}

\end{table}

In a second step, we implemented an open source software, which captures
the incoming data-stream from the sensors, partitions the stream into
data-chunks, timestamps each chunk, and finally stores the chunks. The
software and installation manual, as well as links to these initial open
science datasets, are available from our GitHub repository. To install
the software, the Raspberry Pi needs to be connected to a computer.
During operation the hardware needs power and an internet connection.
The software reads the data-stream from the sensors and appends it to an
internal buffer. After a user-defined time, or when the data volume size
exceeds a chunk size of about 256 kB, a new chunk is created. A hash of
this data chunk is calculated and uploaded to the trusted timestamping
service Originstamp\footnote{www.originstamp.org}, which stores it on
Bitcoin's blockchain {[}10{]}. Therefore, a personally owned API key is
required. That key can be used to identify the creator of the dataset,
thus signing the data itself is not required. Beginning from the second
chunk, a reference to the previous chunk is included in the hash.

However, the hash is useless without the data. Many services exist to
upload data to a central server. However, this central resource might
fail or get compromised. We therefore use the Interplanetary Filesystem
(IPFS) to be independent of a central authority for storage. IPFS itself
is organized in blocks and the block-address is the hash of the file
content, which we already used to generate the timestamp. Therefore, we
installed IPFS on the Raspberry to upload the data chunk by chunk.

For long-time archival, we plan to copy the finalized measurement
timeseries from IPFS to the long-time archival platform Zenodo\footnote{www.zenodo.org}.

\section{Discussion and Future Challenges}

In section 3, we demonstrated how to guarantee the immutability of time
series of measurement values with inexpensive hardware (challenge II).
While this line of research has many open questions when the data volume
of the incoming data stream of sensor data increases, a first step has
been completed. We argued before (Section 1) that a highly distributed
network of citizens performing measure reduces the chance for
intentional data fabrication or malicious device tampering. One model
use case, such as temperature recording for climate research has easy
achievable requirements, with regards to the 5 requirements for
trustworthy data (challenge I). The instruction of the participants and
the development of algorithms to automatically detect measurement time
series that do not fulfil the requirements requires extra effort.
Especially the collection of meta data, for instance the location of the
experiment is hard to verify automatically, even though first approaches
for verifying location exist {[}2{]}. Proof-of location is based on
token incentivized, community driven data. People witness their mutual
locations supported by hardware IDs from network cards and Bluetooth
controllers to provide evidence that a certain device was at a certain
point.

Moreover, for many use-cases the required error rate is too low, the
sampling rate too high or the overall complexity of the experiment is
too high so that citizens must be replaced by a smaller number or
institutions that have the capabilities to perform the experiments.
While the same methods can be applied to ensure the immutability and
durability as outlined before, the chances for tampering and
unintentional mistakes during measurement remains. While trusted
timestamping can guarantee that data has at least a certain age this
does not prevent one from recording data without timestamping it,
manipulate the data thereafter pretend that the manipulated data is
current. This is different from the physical world, where the age of an
object can be determined using forensic methods. As a result, we
currently are unaware of any data-driven approach to ensure the
trustworthiness of an individual data source. Thus, if only one, or a
few data sources exist the classical author reputation is the only
method to estimate the data quality.

\section{Conclusion}

Measurement values and data streams from sensors are not immune to
manipulation or retrospective selective pruning. The ability to securely
prove that data was not manipulated or omitted is especially important
for both open science and citizen science projects. In this paper, we
proposed a method for independently and securely verifying the time of
creation of sensor data, as well as tracing it to its rightful owner. By
storing the data in a decentralized manner on the Interplanetary
Filesystem and by relying on a blockchain-backed solution for storing
tamper-proof and decentralized trusted timestamps associated with
discrete chunks of measurement values, we showed how sensor data can be
made securely verifiable. Our exemplary Raspberry Pi prototype
demonstrates how our proposed modifications can be easily implemented
into commonly used hardware sensors, such as temperature, electric
current, and vibration detection sensors. We argue that enabling any
researcher or interested citizen to verifiably trace the integrity of
measurement values and their time of origin can significantly strengthen
the open science movement and can increase the trustworthiness of
citizen science projects.

\section{References}

{[}1{]} Beck, R., Czepluch, J.S., Lollike, N., and Malone, S.
Blockchain-the Gateway to Trust-Free Cryptographic Transactions.
\emph{ECIS}, (2016).

{[}2{]} Brambilla, G., Amoretti, M., and Zanichelli, F. Using Blockchain
for Peer-to-Peer Proof-of-Location. \emph{arXiv preprint
arXiv:1607.00174}, (2016).

{[}3{]} Casino, F., Dasaklis, T.K., and Patsakis, C. A systematic
literature review of blockchain-based applications: current status,
classification and open issues. \emph{Telematics and Informatics},
(2018).

{[}4{]} Catalini, C. and Gans, J.S. \emph{Some simple economics of the
blockchain}. Rotman School of Management Working Paper No. 2874598; MIT Sloan Research Paper No. 5191-16.  (2017).\newline
http://dx.doi.org/10.2139/ssrn.2874598 

{[}5{]} Fanelli, D. How many scientists fabricate and falsify research?
A systematic review and meta-analysis of survey data. \emph{PloS one 4},
5 (2009), e5738.

{[}6{]} Fanelli, D. Do pressures to publish increase scientists' bias?
An empirical support from US States Data. \emph{PloS one 5}, 4 (2010),
e10271.

{[}7{]} Gipp, B., Breitinger, C., Meuschke, N., Beel, J., and
Breitinger, C. CryptSubmit: Introducing Securely Timestamped Manuscript
Submission and Peer Review Feedback using the Blockchain.
\emph{Proceedings of the ACM/IEEE-CS Joint Conference on Digital
Libraries (JCDL)}, (2017).

{[}8{]} Gipp, B., Meuschke, N., and Gernandt, A. Decentralized Trusted
Timestamping using the Crypto Currency Bitcoin. \emph{Proceedings of the
iConference 2015}, (2015).

{[}9{]} Grech, A. and Camilleri, A.F. Blockchain in education. \emph{Publications Office of the European Union 2017, 132 S. - JRC Science for Policy Report, Luxembourg } (2017).
\newline
http://dx.doi.org/10.2760/60649

{[}10{]} Hepp, T., Schoenhals, A., Gondek, C., and Gipp, B. OriginStamp:
A blockchain-backed system for decentralized trusted timestamping.
\emph{Information Technology 60}, 5--6 (2018), 273--281.

{[}11{]} Nakamoto, S. Bitcoin: A peer-to-peer electronic cash system.
(2008).
\newline
https://bitcoin.org/bitcoin.pdf

{[}12{]} Nosek, B.A., Alter, G., Banks, G.C., et al. Promoting an open
research culture. \emph{Science 348}, 6242 (2015), 1422--1425.

{[}13{]} Schubotz, M., Breitinger, C., Hepp, T., and Gipp, B.
Repurposing Open Source Tools for Open Science: a Practical Guide. 2018.
\newline
https://doi.org/10.5281/zenodo.2453415.

{[}14{]} Seebacher, S. and Schüritz, R. Blockchain technology as an
enabler of service systems: A structured literature review.
\emph{International Conference on Exploring Services Science}, (2017),
12--23.

{[}15{]} Sharples, M. and Domingue, J. The blockchain and kudos: A
distributed system for educational record, reputation and reward.
\emph{European Conference on Technology Enhanced Learning}, (2016),
490--496.

{[}16{]} Yli-Huumo, J., Ko, D., Choi, S., Park, S., and Smolander, K.
Where is current research on blockchain technology? --- a systematic
review. \emph{PloS one 11}, 10 (2016), e0163477.

{[}17{]} Zheng, Z., Xie, S., Dai, H., Chen, X., and Wang, H. An overview
of blockchain technology: Architecture, consensus, and future trends.
\emph{Big Data (BigData Congress), 2017 IEEE International Congress on},
(2017), 557--564.

\end{document}